\documentclass
[%
superscriptaddress,
preprint,
 amsmath,amssymb,
 aps,
longbibliography]{revtex4-2}

\usepackage{graphicx}
\usepackage{dcolumn}
\usepackage{bm}
\usepackage{comment}

\graphicspath{ {./} }
\usepackage{svg}

\usepackage{xcolor}
\usepackage{wasysym}
\usepackage{fdsymbol}

\usepackage{xcolor} 

\begin{document}

\preprint{Subm. to INTERFACE FOCUS. \hspace{1cm}           IUTAM: Capillarity and Elastocapillarity in Biology special issue.}

\title{Channel deformations during elastocapillary  spreading of gaseous embolisms in biomimetic leaves.}

\author{François-Xavier Gauci}
\affiliation{
Université Côte d’Azur,  CNRS UMR 7010, Institut de Physique de Nice (INPHYNI), 17 Rue Julien Lauprêtre, 06200 Nice, France}
\author{Ludovic Jami}
\affiliation{
Université Côte d’Azur,  CNRS UMR 7010, Institut de Physique de Nice (INPHYNI), 17 Rue Julien Lauprêtre, 06200 Nice, France}
\author{Ludovic Keiser}
\affiliation{
Université Côte d’Azur,  CNRS UMR 7010, Institut de Physique de Nice (INPHYNI), 17 Rue Julien Lauprêtre, 06200 Nice, France}
\author{Celine Cohen}
\affiliation{
Université Côte d’Azur,  CNRS UMR 7010, Institut de Physique de Nice (INPHYNI), 17 Rue Julien Lauprêtre, 06200 Nice, France}
\author{Xavier Noblin}
\affiliation{
Université Côte d’Azur,  CNRS UMR 7010, Institut de Physique de Nice (INPHYNI), 17 Rue Julien Lauprêtre, 06200 Nice, France}
\email{xavier.noblin@univ-cotedazur.fr}

\date{\today}

\begin{abstract}
The nucleation and/or spreading of bubbles in water under tension (due to water evaporation) can be problematic for most plants along the ascending sap network from root to leaves, named xylem. Due to global warming, trees facing drought conditions are particularly threatened by the formation of such air embolisms, which spreads intermittently and hinder the flow of sap and could ultimately result in their demise. PDMS-based biomimetic leaves simulating evapotranspiration have demonstrated that, in a linear configuration, the existence of a slender constriction in the channel allows for the creation of intermittent embolism propagation (as an interaction between the elasticity of the biomimetic leaf (mainly the deformable ceiling of the microchannels) and the capillary forces at the air/water interfaces) \cite{Keiser2022}-\cite{keiser2024}. 
Here we use analog PDMS-based biomimetic leaves in 1d and 2d. To better explore the embolism spreading mechanism, we add to the setup an additional technique, allowing to measure directly the microchannel's ceiling deformation versus time, which corresponds to the pressure variations. We present here such a method that allows to have quantitative insights in the dynamics of embolism spreading. The coupling between channel deformations and the Laplace pressure threshold explains the observed elastocapillary dynamics. 
\end{abstract}

\maketitle

\section{\label{sec:level1} Introduction}

 The nucleation and growth of bubbles in plant can have dramatic (positive or negative impacts). In the process of fern spores’ ejection, the nucleation of cavitation bubble in water under tension triggers the fast catapulting mechanism. The origin of this fast motion lies in the deformation of a catapult-like elastic structure (the annulus) induced by water evaporation. This later leads to the annulus compartments volume decreasing and negative pressure building up, up to the nucleation of a cavitation bubble. This later triggers the fast catapulting mechanism that we have studied in real and artificial systems \cite{Llorens2016}-\cite{Scognamiglio2018}. The same phenomenon of nucleation of cavitation bubbles in water under tension (due to water evaporation) can be problematic for most plants along the ascending sap network from root to leaves, named xylem. 

 In plants, the sap circulates from roots to leaves through the veins of their xylem system. A vein is a bundle of vessels made of sequences of hollow vessel elements.
Theoretical explanation for sap flows in trees has been first described in 1895 by the cohesion-tension theory \cite{Dixon1895}. The driving force behind the rise of sap in plants is the evaporation occurring in leaves. It creates a gradient of negative pressure along the water-filled xylem vessels that pulls the water column upwards \cite{Tyree2013} \cite{Stroock2014} \cite{venturas2017}. It allows the sap to rise several tens of meters from roots to leaves but in return, leads to a risk of emergence of gas embolism, namely the spreading of gas inside the hydraulic network, which can inhibit or even block the sap circulation and seriously impact the productivity of the plant \cite{Choat2016}. This risk of embolism formation is dramatically increased by conditions of water stresses in soils. Global warming and its consequences in terms of increase of drought events frequency and intensity will put most forests across the globe in danger of survival \cite{Choat2018}. 

Mechanisms at the origin of gas embolism in xylem vessels are still not clearly established and could be various. A tension too high can generate cavitation events \cite{Wagner2022} or it may come from pre-existing gas bubbles associated with hydrophobic vessel walls \cite{Schenk2017} or air-seeding, coming from outside. 
 
 For leaves, the occurrence of embolisms could be observed directly to spread intermittently and possibly result in catastrophic events \cite{Brodribb2016PNAS}. PDMS-based biomimetic leaves simulating evapotranspiration has demonstrated that, in a linear configuration, the existence of a slender constriction in the channel allows for the creation of intermittent embolism propagation (as an interaction between the elasticity of the biomimetic leaf (mainly the deformable ceiling of the microchannels) and the capillary forces at the air/water interfaces) \cite{Keiser2022}-\cite{keiser2024}.

\begin{figure}[h]
   \includegraphics[width=1\textwidth]{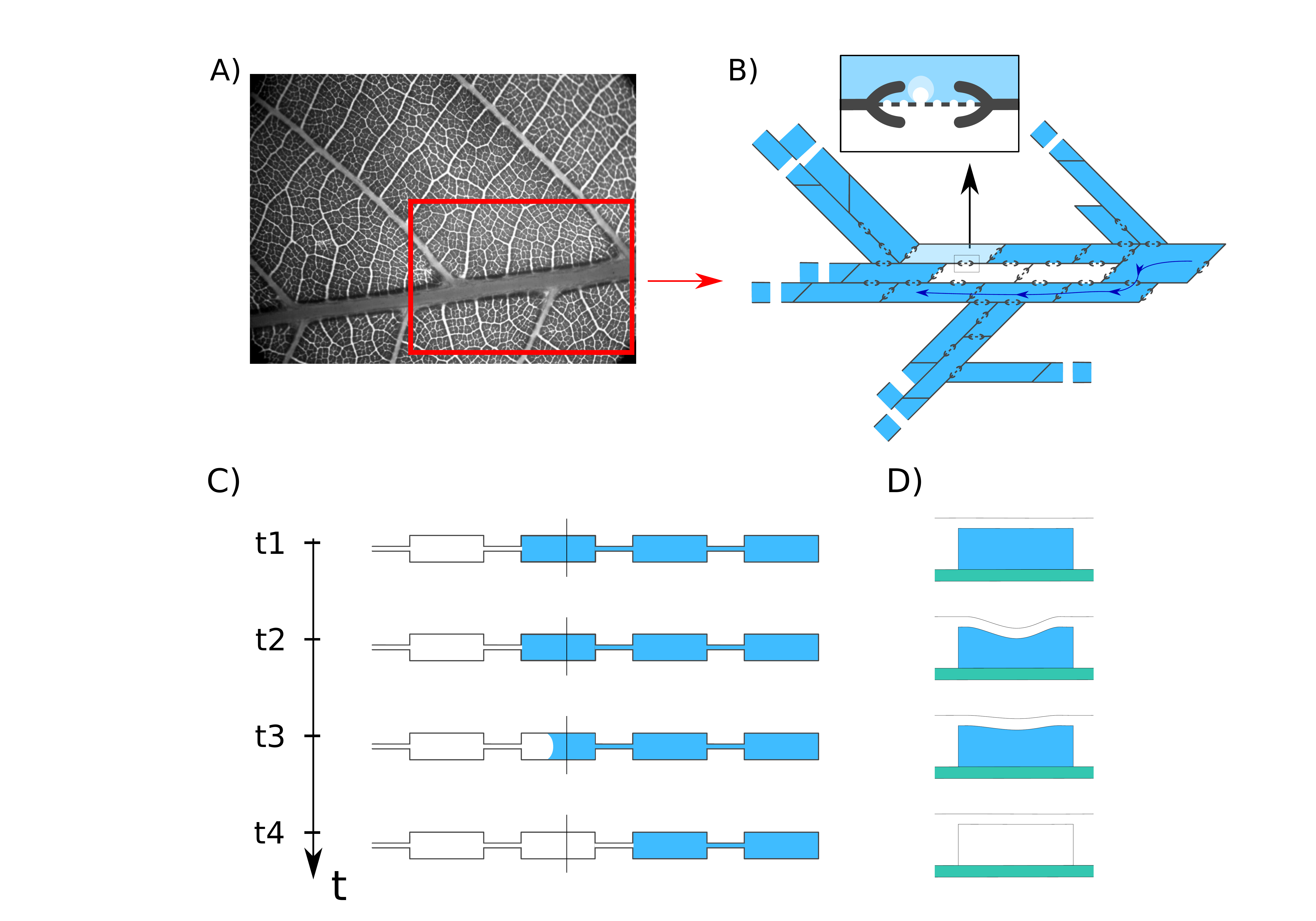}
    \caption{A) Image of a leaf of Walnut tree (Tahnks to Virgile Thiévenaz). B) Scheme of xylem vessel elements connected by bordered pits with embolism spreading. C) Top view of biomimetic systems with identical channels connected by narrower constrictions at various times. D) Side views (taken at black lines locations) of the deformation of the microchannels ceilings as the embolism passes through it.}
    \label{Fig1}
    \end{figure}

Once inside the system, embolism spreads in vessels elements step by step, intermittently and hierarchically \cite{Brodribb2016PNAS,Brodribb2016NP}. On the one hand, the intermittency is characterized by fast events of gas invasion in vessels, followed by pauses. It comes from the presence of wall pits longitudinally and laterally connecting vessels of the same vein (Fig. \ref{Fig1} B)) \cite{Jensen2016}. These micrometric holes contain cellulosic mesh and play as defense valves that momentarily contain the embolism. On the other hand, hierarchical spreading manifests itself by starting preferentially from the main veins (the midrib and secondary veins) and ending by the thinnest ones (higher order veins).

The three-dimensional structure of a leaf, its variety of materials, size scales and its complex structure make difficult interpreting a transmission view of a drying leaf. A biomimetic approach makes images acquisition simpler, allows to control parameters and environmental conditions more accurately and isolate specific physical ingredients.

Leaves evapotranspiration can be reproduced, using Polydimethylsiloxane (PDMS) thanks to its permeability to gas and water\cite{Noblin2008}, so a micro channel network can be built to create a drying PDMS leaf venation system.

The presence of pits has already been mimicked artificially, by linking two PDMS microchannels by a narrower one \cite{Keiser2022}. When the air / water interface gets into the constriction, its curvature increases strongly, lowering pressure in the exit liquid-filled channel. The non-linear dynamics of the meniscus progress comes from an elasto-capillary coupling between the compliance of the channel and the interface curvature, forced to sustain large changes to get through the constriction. Experiments with several identical channels in series have also been performed \cite{keiser2024} in which a sequence of jumps and stops of the meniscus is observed, due to the presence of constrictions.

In the present work, we develop a new experimental setup to couple visual observation and instantaneous pressure measurement from the microchannel ceiling deformation of analogous biomimetic PDMS-based systems .To measure the instantaneous pressure of water inside a microchannel, we developed a new method based on interferometric deformation measurement. We used 1d systems with successive identical channels similar to the one used in \cite{keiser2024}, as shown in Fig. \ref{Fig1} C) D). Then we have explored, for the first time embolism propagation in a 2d system that is closer to real leaves. 

In this manuscript we present first the microfluidic devices used. We then explain the interferometric deformation measurement method used along with the calibration of such measurement device to measure pressure variations. We then present results obtained for the pressure variations versus time in both type of systems. We then discuss these findings and develop a conclusion with several perspectives. 

\textcolor{red}{}

\section{Materials and Methods}

\subsection{The biomimetic pervaporation leaf setup}
\subsubsection{Biomimetic leaf.}
We used two types of devices: 1) in 1d, the same as used in \cite{keiser2024} where identical channels elements are connected by identical constriction much narrower than the channels. The depth and the width of the constrictions are respectively  $48 \,\mu m$ and $ 22\,\mu m$ while the channels have the same depth but a width of $303 \,\mu m$  (Fig. \ref{Fig1} C).

 2) In 2d, we used a new device composed of interconnected microchannels presenting between one and four neighbors of different widths connected by a constriction of depth $38 \,\mu m$ and width $ 20\,\mu m$. The scheme is presented in Fig \ref{Fig5}.
 
 The PDMS devices have been made following a classical protocol of soft-lithography \cite{mcdonald2000fabrication}. To do so, we first designed the photolithography masks using a Python program which returns a svg file that has been printed on a plastic film by the Selba company. Then, we created the network mold in SU8 negative resist (MicroChem) from which we made the final PDMS device.
To ensure a reduced thickness of the ceiling above the channels, the mixture of liquid polydimethylsiloxane (PDMS; SylgardTM 184, from Dow company) and curing agent (in the mass fraction 10:1), is spincoated on the SU8 mold. The liquid PDMS mixture in the mould is then degassed under vacuum and heated in a oven at 65°C during 24 hours to obtain a thin sheet of PDMS that we will call the PDMS membrane in the following. 

Once the PDMS membrane is reticulated, it is removed from the mould and put on a plasma-treated glass slide. With a scalpel, we perform a wide cut at the petiole end (which is the base of a leaf network connected to the rest of the tree, for us, it is the starting point for the air in our device), creating a direct connection between the network entrance and the outside.  The system is then immersed in water in an open Petri dish and placed under vacuum inside a desiccator during several hours. The air leaves the microchannels which fill with water coming from the cut. We obtained the biomimetic leaf which will be used for the embolism experiment.
\subsubsection{Pervaporation setup.}
The imaging setup is composed as follows. A white light source is positioned underneath the biomimetic leaf and we image it by reflection thanks to a semi-reflective glass so the air-filled channels appear bright on the camera (PixeLink). As the PDMS is permeable to water and the leaf is exposed to a dry circulating air flow, the water in the channels pervaporates through the PDMS membrane. As soon as the device is out of water, it starts pervaporating and as the time needed to prepare the experiment is fluctuating, we define the reference time as the moment the first channel gets embolized. The air gradually invades the system from the cut at the petiole. The embolism spreads through the  network, air replacing water.

We capture a picture of the embolizing leaf every 10 seconds and make a time-lapse for the duration of the experiment which is about 5 hours. The result is a stack of .tif images. 

We also record simultaneously the deformation of a choosen channel ceiling, as described in the next paragraph.

\begin{figure}[h]
\includegraphics[width=1\textwidth]{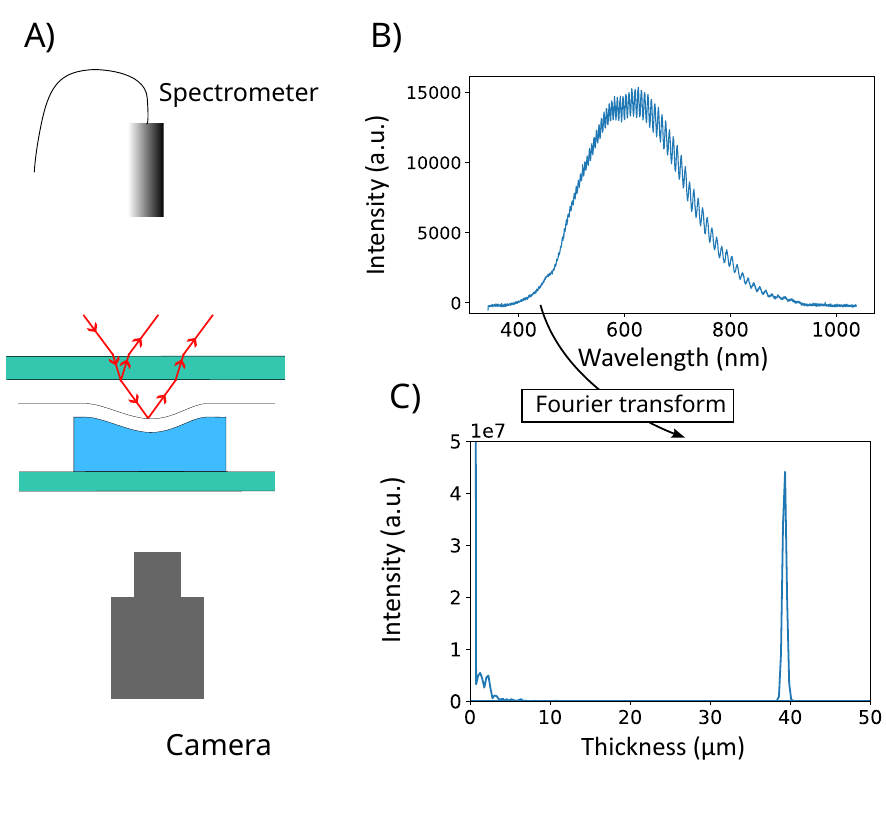}
\caption{A) Setup with camera and spectrometer to record spectra for thickness measurements. B) Spectrum showing oscillations due to the air thicness between glass slide and ceiling of microchannel. C) Fourier trasnform of the spectrum giving a peak at the thickness measured.}
\label{Fig2}
\end{figure}

\subsection{Deformation measurement.}

\subsubsection{Experimental setup.}

The setup we developed to measure the pressure in the channel during the embolization of the device is schematized on figure \ref{Fig2}. The pressure is calculated from the resulting deformations of the channel ceiling under the effect of the depression or repressurization in the channel during embolism. To measure the deflection of the ceiling, we used an interferometry-based method.
However, as Water and PDMS optical indices are too close to use directly interferometry, we created an extra air layer by adding a glass slide some tenth of micrometers over the system (Fig \ref{Fig2}). Dry air is circulated in this space to ensure that the device dries. We obtain the deflection of the membrane by measuring the thickness of this air layer over time.

During the experiment, the spectrometer is positioned directly above a channel and supplies the wavelength spectrum to a program that computes the air thickness every two seconds by a Fourier transform which then gives the channel deformation $\zeta$. 
Deformation at rest is measured when the channel is fully embolized, so the air inside is assumed to be at the pressure of the lab atmosphere.

\subsubsection{Calibration}

\begin{figure}[h]
\includegraphics[width=0.8\textwidth]{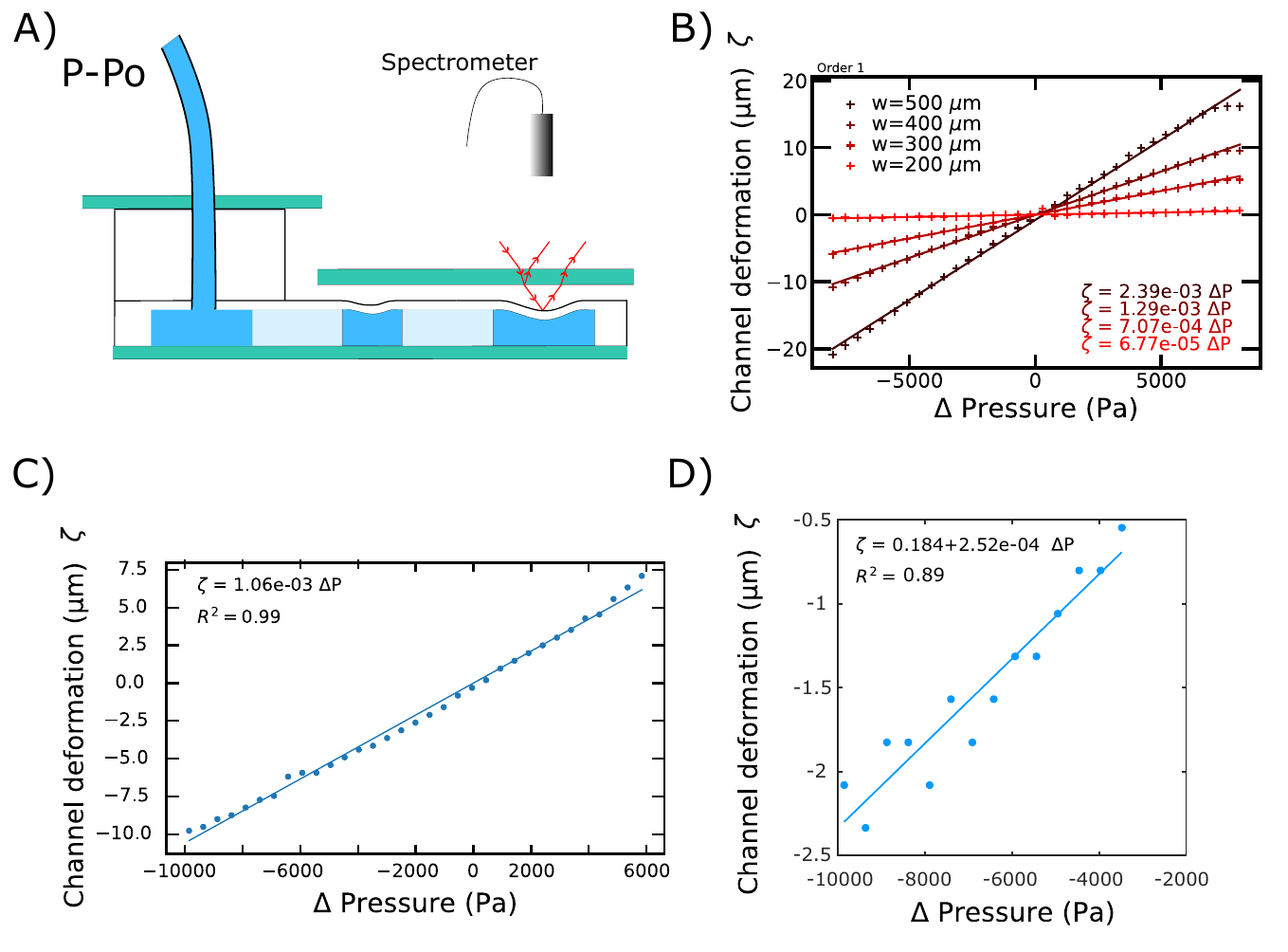}
\caption{A) Setup for the calibration of pressure / deformation curves. B) Calibration curve on 2D systems similar to the ones used in experiments, for which different channel widths have been tested. C) Calibration curve obtained a posteriori on a channel after an experiment was done, in a 2d system and in D) for a 1d system.}
\label{Fig3}
\end{figure}

We performed calibration experiments for a channel element of a given width and a given ceiling thickness, to relate the deformation of the channel ceiling to the pressure in the channel element.
More precisely, we impose the pressure (by gravity) in the channel and simultaneously measure the resulting deformation of the membrane with the spectrometer. We obtained a linear relation between the deformation of the membrane and the pressure in the channel element (Fig. \ref{Fig3}). So, for a channel element of a given width and a given ceiling thickness, we can deduce the pressure in the channel by measuring the deformation of its ceiling.
We made this calibration for the different channels of the bio-mimetic leaf (Fg. \ref{Fig3}). 

This calibration has been done following two methods: 1) on similar devices as the one used in the experiments, for which the inlet for pressurized water was bonded by plasma (but then not used in evaporation experiments). 2) on the exact same devices that were used for experiments, with a scar at the inlet over which a pdms piece was pressed on. This a posteriori calibration allows us to obtain it over the same exact device used for experiments. A perforated PDMS block is placed at the level of the cut in the petiole and maintained with clamps to ensure impermeability. The device is then filled with water and connected to a syringe via a tube. We then vary the altitude of the syringe and can compute the channel deformation for different pressures.

\newpage

    \section{Results}

\subsection{1d system}

\begin{figure}[htbp]
\includegraphics[width=1\textwidth]{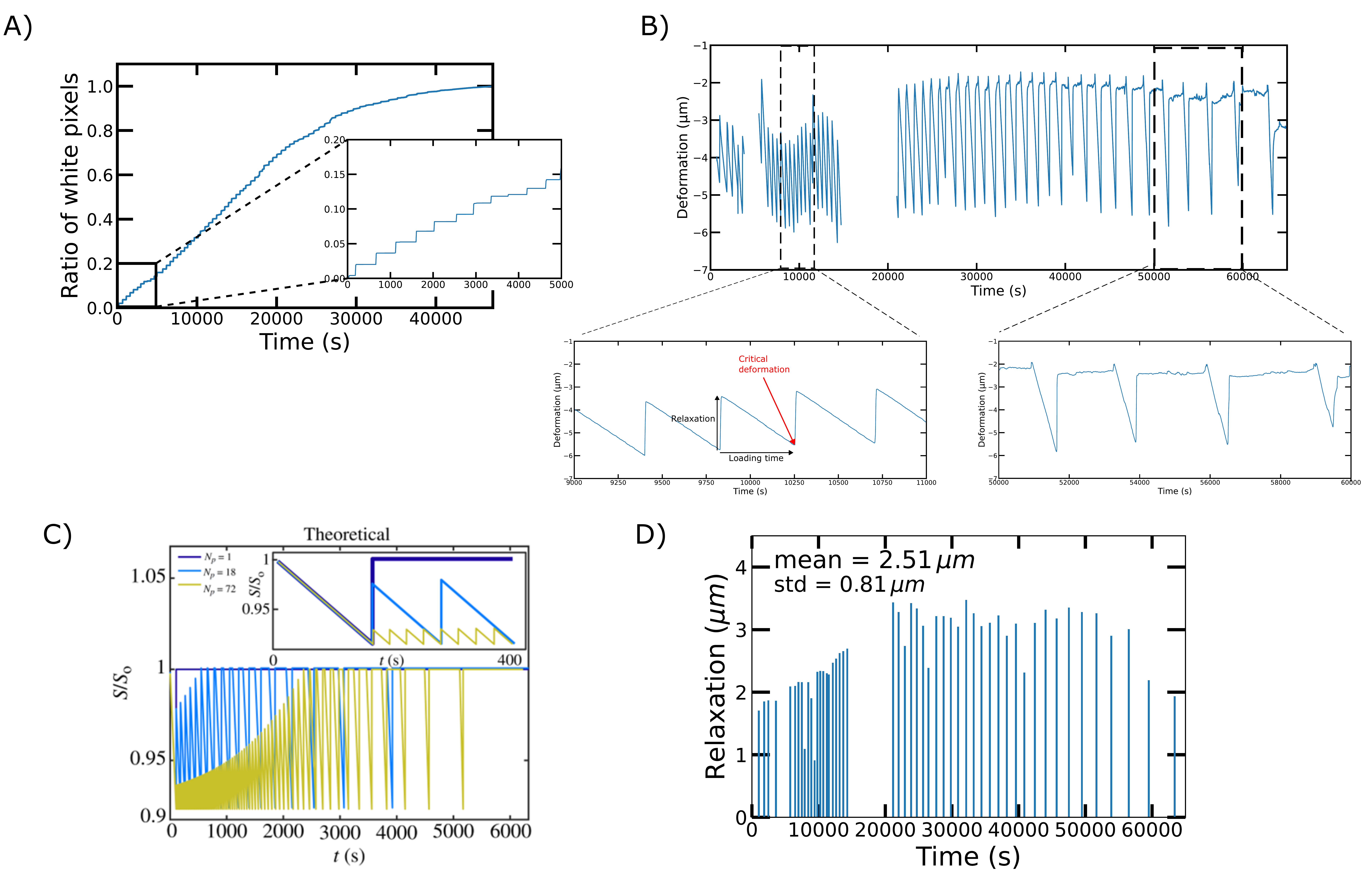}
\caption{ A) Ratio of detected white pixels in a 1D system vs. time. B) Deformation as function of time for this 1d system. C) Theoretical curves predicted in \cite{keiser2024} for the section area of channel. D) Jumps amplitude vs. time for the deformation in the measurement channel vs. time.}
\label{Fig4}
\end{figure}

In Fig. \ref{Fig4}A), we plot the amount of embolized pixels over time, this expresses the surface area of the gaseous phase as function of time divided by the final total area of the systems entirely filled by air.

 Globally, the embolism rate slightly slows down over time (FIG. \ref{Fig4}), because the evaporation flux is proportional to the wetted surface of the system \cite{Noblin2008, Dollet2019}.
By zooming, we observe an intermittent dynamics, sequences of arrests and jumps of the bubble, similar to the experiment of Keiser et al. \cite{keiser2024}.

From the deformation setup, we estimate the channel maximal deformation at its center and plot it over time. The curve in Fig.\ref{Fig4} B) presents the deformation curve of the last channel element of the system.

By zooming (two inserts below), the graph for early time ($t<\, 20000 s$) shows an alternating gradual declines (during pervaporation and total volume decrease of water in all the system) and sharp increase when the meniscus unpins. This means a succession of depressurization and sudden relaxation of the targeted (measurement) channel. This repeats each time the embolism passes a constriction. 
The cycle frequency decreases throughout the experiment whereas the amplitude of the jumps increases over time. During this first part of the experiment, the jumps are complete (and pressure relaxation incomplete) and correspond to going directly from t2 to t4.

For the second part of the experiment ($t>20000 \, s$), as seen on the second insert (right graph), the amplitude of the jumps are more or less constant and the frequency is still decreasing. Instead of presenting a slow decrease and sharp increase, it passes then by a plateau phase since when the meniscus jumps, it does not fully advances in the channel element. As drawn in Fig.\ref{Fig1}C), it corresponds to the phase at time t3. During all the moment where the meniscus displaces inside a channel element, the deformation stays constant. The jumps are incomplete, but the pressure relaxations are complete.

The waiting time discussion is detailed in \cite{keiser2024}. Here we are in a situation that corresponds to a large number of channel elements $N_p$ and which is similar as predicted theoretically in \cite{keiser2024}. We have chosen in B) to plot the deformation from our results. In C), the relative surface variation of the wetted section of the channel is plotted from this article. In the small deformations regime, this is completely equivalent since we have $S=h*w$ and by noting $h(t)=h_0-\zeta(t)$ we have $S=S_0-0.5 \zeta*w$ and $S/S_0=1-0.5 \zeta/h_0$. 

In D) we have plotted the amplitude of each jumps (fast relaxation in pressure as noted in the insert). This allows an easy comparison with the curve from the theoretical approach from \cite{keiser2024} in C). We observe the same trends as discussed before on the increase in jumps amplitude, reaching a plateau. The data are interestingly noisy, even in this device in 1d with all constrictions and channels being identical.

\subsubsection{Comparison between theoretical and experimental amplitude of pressures jumps.}

The average value for the jumps is around 2.5 microns. The plateau value is close to 3 microns. Using the calibration made for the 1d system, we can determine the corresponding pressure differences. 

The theoretical pressure differences at which the meniscus passes a constriction are given by the Laplace law and for a rectangular section being $h_c=48\,\mu m$ and $w_c=22\,\mu m$. It writes $\Delta P_c=\gamma \kappa$, with :

\begin{equation}
   \kappa = \frac{\left(1+ \frac{w_c}{h_c}\right)+\sqrt{\left(1-\frac{w_c}{h_c}\right)^2+\pi\left(\frac{w_c}{h_c}\right)}}{w_c}
\label{eq:kappa}
\end{equation}

Here, the threshold given by the Laplace pressure is then equal to: $\Delta P_c(theo)=9079 Pa$, assuming no effect of contact angle.

We can compare this value to the plateau value Fig.\ref{Fig4} D which is around $3\, \mu m$. In terms of pressure, using the slope obtained in calibration in Fig.\ref{Fig3} D), this corresponds to $ \Delta P_c(exp)=11905 Pa$. This experimental value of the pressure jump is then slightly higher, but in very good agreement with the theoretical value $P_c(theo)=9079 Pa$.

The maximal possible curvature compatible with the geometrical constraints only depends on the size of the constriction (the most important is the width when smaller than the thickness, but this latter is also important). The thinner the constriction, the greater the pressure difference required for the meniscus to pass through it. 

The length of a meniscus jump depends directly on the amount of water lost by the system during the loading time. At the beginning of the experiment with still a large water reservoir, a great amount of water pervaporates during the loading time. By mass conservation, the potential length of the meniscus jump is proportional to this volume. Depending on the size of the exit channel, this jump can be limited and the new front directly pinned to the next constrictions. In that case, the depression in the liquid volume is not completely released, and a new loading starts directly after the jump. If the exit channel volume is large enough compared to the potential length of the meniscus jump, the meniscus stops inside the exit channel.

\subsubsection{Comparison between hydrodynamic and capillary pressures.}

Between two neighbouring channels, the constriction resistance is $R_h=\frac{12\eta L_c}{(1-0.63\frac{w_c}{h_c})h_cw_c^3}$ for a rectangular channel and the flux is about $Q_{perv}=q_{lin}\frac{w}{\delta}L$ with $q_{lin}=D_pC_p^{sat}(1-RH)$ \cite{Keiser2022}. So the pressure gradient due to hydrodynamic resistance is $\Delta P_{hydro}=R_hQ_{perv}\sim 3.10^{-1} Pa$.
On the other hand, the pressure gradient due to capillarity is about $\Delta P_{cap}=\frac{\gamma}{w_c}\sim 7.10^{3} Pa$.

The constriction length is $L_c=230 \mu m$, width: $w_c=22 \mu m$, height; $h_c=48,3 \mu m$. we can deduce the hydrodynamic resistance $R_h=7,53.10^{12} Pa.s.m^{-3}$

$\Delta P_{cap} >> \Delta P_{hydro} $ leads us to assume that the pressure is uniform in the liquid volume at a given time. 

\subsection{2d system}

\begin{figure}[htbp]
\includegraphics[width=0.9\textwidth]{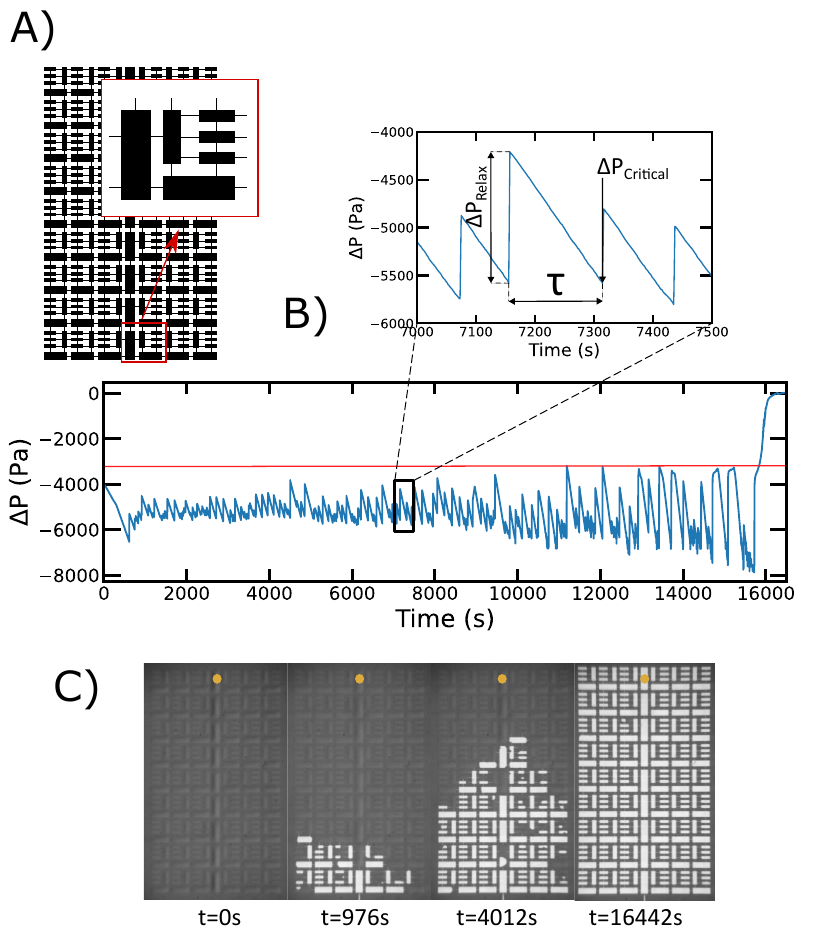}
\caption{A) Scheme of 2D device with various channel sizes. B) evolution with time of the ceiling deformation of the channel pointed in C). C) Embolism spreading with time. White domain represent the spreading air embolism. The yellow dot points the channel element where the deformation is measured over time. It correspond to the largest channels element, at the end of the main vein, the most far away from petiole (start of the embolism).}
\label{Fig5}
\end{figure}

In the 2d case, during the drying, the embolism front is made of several menisci in channels or constrictions. Once they are all pinned at constrictions, water-filled channels still pervaporate, lowering the pressure and shrinking channels until the pressure reaches a low enough value locally. After that, one meniscus jumps into its exit channel, suddenly increasing the pressure and relaxing the channel cross section. The leaf is made of a pattern of water or air connected domains; hence, clusters form that can isolate water connected domains. Each of these clusters are then independent on a hydrodynamic point of view. This can be seen, for example, in Fig. \ref{Fig5} C). for t=4012s. 

The deformation curve is much less regular than in the 1d case. We have used the calibration curve to plot here directly the pressure, in (Fig.\ref{Fig5} B)). The jumps are globally increasing in amplitude on average, as in the 1d case, but they  show a larger variability and randomness for their height and the waiting time between two relaxations. 

Here, the width of the constriction is $w_c=20 \, \mu m$ and height $h_c=38\, \mu m$. The theoretical critical pressure for the full jump is then, following Eq. \ref{eq:kappa},  $\Delta P_c(theo)=10426 Pa$. In the experiment, the critical pressure for each passed constriction is given by the lowest value of each slow pressure decrease curve, as shown in insert Fig.\ref{Fig5}. The values lie between 6000 Pa and 8000 Pa which is a bit lower than the expected theoretical critical pressure of $\Delta P_c(theo)=10426 Pa$. It is normal that for not complete pressure relaxation, the jump are lower.

\section{Discussion}

For 1d systems, A very good agreement is found for the deformation curves as function of time that are fully analogous to the theoretical curves predicted in \cite{keiser2024} for various aspects: Jumps frequencies, amplitudes, waiting time and complete and non-complete relaxation. Two main slight discrepancies can be noted: 1)  The critical pressure at which the jump occurs varies for each jump. Obviously in experiments with real systems, the geometry of each constriction cannot be  strictly identical. The main explanation may be that the microfabrication is not perfect and the constriction sizes is not constant throughout the network, giving a random aspect to the problem. 2) The calibration to obtain absolute value of pressure from deformation could be improved. Comparing the $R$ parameter of both fits in Fig \ref{Fig3} (0.99 for 2d then 0.89 for 1d) shows that for stiff channels (thick Top PDMS Layer / narrow channels...), the precision is lower, even if the value found is quite close. The important point to note in our study is that relative measurements throughout all the experiment duration are very precise. It is only the comparison of experimental and theoretical absolute values that can be less precise.

The close results between the volume conservation-based model and experiments are compatible with our hypothesis of a homogeneous pressure throughout the system. As calculated above, the hydrodynamic pressure is much less than the capillary pressure, leading to this uniform value.

For 2d systems, that are tried for the first time, a more complex dynamics with higher randomness in the pressure jumps amplitude, waiting times and amplitudes distribution, along with spatial clusters formation. Even if the constriction sizes are the same throughout the systems, this could impose a regular pattern as in 1d, the network topology and geometry induces a more complex dynamics.

\section{Conclusion}

1. We present a new technique to measure the pressure variation in the elastocapillary propagation of gaseous embolism in biomimetic leaves.

2. The value of pressure jump have been measured as function of time and are in good agreement with the capillary Laplace law written with the size of the constrictions.

3. The 2d case shows a more complex dynamics.

4. More detailed studies are needed to make the pressure measurements more precise and even if this is improved, a better understanding may be needed to explain the discrepancies with the theoretical pressure jump amplitude that may remains.

\textbf{Perspectives.} Now that a method have been developed to measure the pressure variation, it makes this system a promising and versatile tool to be used for addressing outstanding questions related to embolism dynamics. 

1. 2d experiments are more variable, with pressure curves more complex, this will be studied in a future work by changing constriction sizes and channel connectivity between various channel type. The concept of vein order and hierarchy will then be studied.
In our system, the role of hydrodynamic pressure due to viscous losses is negligible. Everything is dominated by capillarity. Interesting and potentially different propagation patterns could be observed through an increase of the hydraulic resistances which would make the hydrodynamic pressure loss across the network non-negligible (larger networks or narrower constrictions). This is also the topic of a future study.

\section{Acknowledgments}

We would like to thank the ANR (project PHYSAP, ANR-19-CE30-0010, IDEX UCA JEDI ANR 15-IDEX-0001) and Université Côte d'Azur for funding. We thank Yaroslava Izmaylova for her help in the device fabrication. We thank Philippe Marmottant and Benjamin Dollet for fruitful discussions. We thank Virgile Thiévenaz for the walnut tree image. We thank JRSI for the image taken from \cite{keiser2024} in Fig. \ref{Fig4}.


\section{References}

\bibliography{biblio_arxiv.bib}
\bibliographystyle{apsrev4-1}

\end{document}